\documentclass[conference]{IEEEtran}
\usepackage{float}
\usepackage{amsmath,amssymb}
\usepackage{cite}
\usepackage{amsmath,amssymb,amsfonts}
\usepackage{algorithmic}
\usepackage{graphicx}
\usepackage{textcomp}
\usepackage{xcolor}
\usepackage{algorithm}
\usepackage{algorithmic}
\usepackage{graphicx}
\usepackage{multirow}
\usepackage{amssymb}
\usepackage{array}
\usepackage[utf8]{inputenc}
\usepackage{cite}        
\usepackage{url}  
\usepackage{paralist}
\usepackage{tabularx}
\newtheorem{mydef}{Definition}

\newcommand{\EnsureOne}{\renewcommand{\algorithmicensure}{\textbf{Output}}}

\def\BibTeX{{\rm B\kern-.05em{\sc i\kern-.025em b}\kern-.08em
    T\kern-.1667em\lower.7ex\hbox{E}\kern-.125emX}}
\begin{document}

\title{\ DRsam:\underline{D}etection of Fault-Based Microarchitectural Side-Channel Attacks in \underline{R}ISC-V Using \underline{S}tatistical Preprocessing and \underline{A}ssociation Rule \underline{M}ining\
\thanks{Identify applicable funding agency here. If none, delete this.}
}


\author{
Muhammad Hassan\IEEEauthorrefmark{1}, 
Maria Mushtaq\IEEEauthorrefmark{2}, 
Jaan Raik\IEEEauthorrefmark{1}, 
Tara Ghasempouri\IEEEauthorrefmark{1} \\

{\IEEEauthorrefmark{1}Department of Computer Systems, Tallinn University of Technology, Tallinn, Estonia} \\
{\IEEEauthorrefmark{2} Télécom Paris, Institut Polytechnique de Paris, Palaiseau, France} \\

Muhammad.Hassan@taltech.ee, Maria.Mushtaq@telecom-paris.fr, \{Jaan.Raik, Tara.Ghasempouri\}@taltech.ee
}

%

\maketitle

\begin{abstract}
RISC-V processors are becoming ubiquitous in critical applications, but their susceptibility to microarchitectural side-channel attacks is a serious concern.  Detection of microarchitectural attacks in RISC-V is an emerging research topic that is relatively underexplored, compared to x86 and ARM. The first line of work to detect flush+fault-based microarchitectural attacks in RISC-V leverages Machine Learning (ML) models, yet it leaves several practical aspects that need further investigation. To address overlooked issues, we leveraged gem5 and propose a new detection method combining statistical preprocessing and association rule mining having reconfiguration capabilities to generalize the detection method for any microarchitectural attack. The performance comparison with state-of-the-art reveals that the proposed detection method achieves up to 5.15\% increase in accuracy, 7\% rise in precision, and 3.91\% improvement in recall under the cryptographic, computational, and memory-intensive workloads alongside its flexibility to detect new variant of flush+fault attack. Moreover, as the attack detection relies on association rules, their human-interpretable nature provides deep insight to understand microarchitectural behavior during the execution of attack and benign applications.

\end{abstract}


\section{Introduction}
Microarchitectural attacks are non-invasive software-based side-channel attacks where the victim’s sensitive information can inadvertently be revealed as a result of performance optimizations, memory access patterns, and timing variations. These performance optimizations on account of compromised microarchitectural security has raised serious concerns about user’s security and privacy~\cite{1,2,3}.
Alongside the continuously evolving microarchitectural attack surface, the concurrent development of advanced detection and mitigation techniques was always there as the first line of defense~\cite{GHASEMPOURI2021114085}. However, despite having a significant number of established solutions, researchers addressed the key concerns of simultaneously ensuring high accuracy and detection speed with minimal performance overhead under the strong threat model~\cite{yuval}. 
In addition, these microarchitectural attacks and their countermeasures are well explored on x86~\cite{5} and ARM~\cite{6} CPUs. Nevertheless, as  RISC-V processors are becoming ubiquitous in critical applications, researchers have also demonstrated that RISC-V based CPUs are also susceptible to microarchitectural attacks~\cite{RISCV}. To this end,  detection of flush+fault based microarchitectural attacks in RISC-V is first carried out by leveraging gem5 and Machine Learning~\cite{mah1,mah3}. The proposed detection method only considers branch mispredictions and instruction cache misses for ML model training and testing. However, relying only on two microarchitectural features considerably increases the chances of higher false positives under the computational and memory intensive workloads. In addition, the ability of the trained Machine Learning model to detect new variants still remains addressable. To address these limitations, we proposed a robust detection method that ensures flexibility to detect new variant of flush+fault attack without any compromise on detection performance even under the stringent conditions. Moreover, the limited interpretability of existing solutions, along with unaddressed research questions, motivated us to propose a white-box human-interpretable detection method for flush+fault attacks in RISC-V, which simultaneously ensures flexibility and reconfigurability.\\ 
To the best of our knowledge, this is the first time that a novel method is proposed for the detection of microarchitectural attacks that combines statistical preprocessing and association rule mining.
\begin{enumerate} 
\item The performance of the proposed detection method is evaluated against fault and return-based variants of flush+fault attack in RISC-V \cite{RISCV} under the stringent conditions, where it outperforms existing solutions under the cryptographic, computational, and memory-intensive workloads.
\item The proposed detection method is more robust and flexible, as it generalizes better to new variants of flush+fault attack, unlike state-of-the-art detection methods that are tailored to a specific type of microarchitectural attacks.
\item The proposed detection method is highly interpretable
contrary to Machine Learning methods, because the attack detection relies on human-interpretable association rules, which provide deep insights to understand attack behavior without compromising detection performance.
\end{enumerate}
The rest of this paper is organized as follows. Section II defines the key terminologies. Section III details the Threat model, experimental setup, and methodology. Section V reports the results of the proposed detection method and its comparison with state-of-the-art, and Section V concludes the paper.

\section{Preliminaries} \label{sec:preliminaries}
In this section, we will provide concise definitions and explanation of terminologies used in this paper.

\begin{mydef} 
\label{DEF1:Raw Data} \emph{\textbf{Raw data}}
The output of gem5 during the execution of attack/benign applications comprises millions of traces against thousands of microarchitectural features. This output is termed as raw data because these traces need to be structured first to get some valuable insights. 
\end{mydef}

\begin{mydef} 
\label{DEF2:Structured Data} \emph{\textbf{Structured data}} is constructed by subjecting raw data to preprocessing, where a set of correlated microarchitectural features is identified and structured.
\end{mydef}

\begin{mydef} 
\label{DEF3:ARM}
 \emph{\textbf{Association rule mining (ARM)}} is an implication of the form $X \rightarrow Y$, where
${X}, {Y} \subseteq {W} \quad \text{and} \quad X \cap Y = \emptyset
$. Where $W$ is a universal set comprising microarchitectural features and label/outcome~\cite{reza}.
\end{mydef}

\begin{mydef} 
\label{DEF4:Support}
\emph{\textbf{Support}} defines how frequently the set of microarchitectural features appears in the data set and is defined as $\text{Supp}(X \rightarrow Y) = P(X \cup Y)$, 
where $P(X \cup Y)$ is the joint probability. In this paper, the support threshold is set to be $\theta_s = $ 5\%~\cite{reza}.

\end{mydef}

\begin{mydef} 
\label{DEF5:Confidence}
\emph{\textbf{Confidence}} provides the likelihood that Y occurs given that X is present.
It is defined as conditional probability $\text{Conf}(X \rightarrow Y) = P(Y \mid X)$ and primarily aims to assess the certainty of the association rule. We set the confidence threshold $\theta_c = $ 90\% for the rule generation~\cite{reza}.
\end{mydef}
\emph{\textbf{gem5}} is a computer architecture simulator that provides huge flexibility to configure multiple ISAs, CPU types, memory models, and cache hierarchies in Syscall emulation or Full system mode~\cite{gem5}. It is notably adopted in microarchitectural security research because it provides access to fine-grained microarchitectural information and helps to discover new vulnerabilities and identify attack behaviors.\\
\emph{\textbf{Apriori}} is a data mining algorithm to generate frequent itemsets based on a prescribed support threshold. In this research paper, we customized the Apriori with minimum itemset size $ \phi = q-3  ,\ \text{where} \ q=8$ \cite{reza}.\\
\emph{\textbf{rv8 benchmark}} comprises cryptographic, computational, and memory-intensive workloads (aes, sha512, norx, dhrystone, qsort, primes, miniz) extensively used for checking RISC-V execution correctness and performance evaluation. We incorporated these workloads to mimic the behavior of benign applications.\\ 
\emph{\textbf{Flush+Fault Attack}} is a Flush+Reload style microarchitectural attack and evaluated by researchers on C906 and U74 RISC-V cores. The two variants of Flush+Fault attack are based on `faulting jump' or `immediate return'~\cite{RISCV}. In the Flush+Fault attack, both the attacker and victim share the same code page to target the particular cache line.
In the fault-based variant, the attacker flushes the instruction cache to bring the cache into the controlled state and takes the first time stamp ($t_1$). Next, the attacker jumps to an address that leads to a fault. Subsequently, the attacker handles the fault and records the time stamp ($t_2$). The time difference of fault handling helps the attacker to deduce whether code in the target cache line was executed or not (instructions are cached or not). This step is followed by dummy jumps outside the target cache line to make sure that target cache lines are never fetched mistakenly by speculation. Return-based variant aimed to further reduce the overhead induced by fault handling. In case any targeted cache line contains a `ret instruction', the attacker jumps directly to it. Hence, replacing the faulty jump with a return instruction and ultimately resulting in low overhead measurements.
\section{Methodology}
\begin{figure}[H]
\centering
\includegraphics[width=3.4in, height=3.3in]{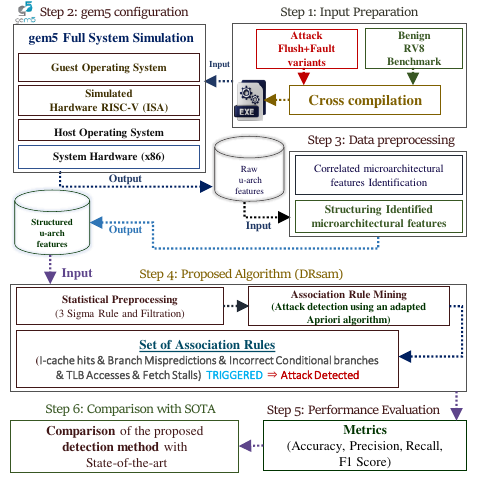}
\caption{Complete workflow of the proposed methodology}
\label{fig:ISCAS}
\end{figure}

In this paper, we propose a robust method for detecting flush+fault-based microarchitectural attacks in RISC-V. In the following, the threat model and the proposed methodology are described.\\
\emph{\textbf{Threat Model}}\\
The proposed method for the detection of flush+fault microarchitectural attacks in RISC-V has the following assumptions: 
\begin{inparaenum}[1)]
    \item Any variant of Flush+Fault attack is running in parallel to cryptographic, computational, and memory-intensive workloads.
    \item Cross-compiled attack and benign executable files are running on gem5.
    \item Operating System (OS) is not vulnerable.
    \item Processor supports speculative execution.
\end{inparaenum}
The entire workflow is divided into 6 steps. Fig.~\ref{fig:ISCAS} illustrates the complete workflow of the proposed methodology.
\\
\textbf{Step 1: Input Preparation}\\
In this step, the source files of flush+fault attack variants and rv8 benchmarks are cross-compiled. The output of this step comprises executable files for RISC-V.
\\
\textbf{Step 2: gem5 configuration}\\ 
This step handles gem5 configuration with RISC-V ISA in Full-System mode. To analyze the execution behavior at the microarchitectural level, the cross-compiled executable files are subjected to gem5. The output of step 2 is raw data (Definition~\ref{DEF1:Raw Data}) comprising millions of traces against thousands of microarchitectural features. Table~\ref{tab:Exp-Setup} (row 2) discusses the gem5 (refer to Section II) configuration setup for detecting flush+fault based microarchitectural attacks.\\
\textbf{Step 3: Data Preprocessing}\\
This step is responsible for data preprocessing. In this step, a set of correlated microarchitectural features affected during the execution of the flush+fault attack is identified and structured to get valuable insights from the gem5 output.  Table \ref{tab:Exp-Setup} (row 3) describes this set of microarchitectural features. In the following, we demonstrate the hypothesis behind collecting  this set. These microarchitectural features are selected due to their relevance and distinct roles in detecting both variants of flush+fault attacks. During the execution of the flush+fault attack, the attacker flushes the instruction cache to bring the cache into a controlled state. Therefore, the next execution instruction needs to be fetched from memory, leading to higher instruction cache misses. In addition, the faulting jumps leading to page faults or jumps to return, alongside thousands of dummy calls, significantly increase branch mispredictions. This significant increase in branch mispredictions is a result
\vspace{-1em}
of incorrectly predicted conditional branches. The extensive number of total executed branches is also linked to the deliberate faulting jumps/immediate returns. Moreover, as a result of instruction cache flushing, the CPU needs to wait unless the instructions are fetched from memory; therefore, the CPU pipeline can’t immediately fetch the next instruction, which yields a higher number of fetch stalls. In the fault-based variant, fetch stalls are even higher than in the return-based variant because page faults induce more stalls in the instruction fetch stage of the CPU pipeline. Finally, the attack execution significantly increases exception handling, memory management operations, stack handling, and kernel activity, which results in a noticeable increase in load and store instructions. Given the preceding analysis, all these discussed correlated microarchitectural features are incorporated for the detection of flush+fault attacks. The output of this step is structured data (Definition~\ref{DEF2:Structured Data}) ready for subsequent stages.
\\
\textbf{Step 4: Proposed Algorithm (DRsam)}\\
This step briefly summarizes the discussion about the proposed algorithm DRsam. The proposed algorithm combines statistical preprocessing and association rule mining. Statistical preprocessing independently flags each microarchitectural feature that exceeds $3$ standard deviation from the mean, whereas association rule mining (Definition~\ref{DEF3:ARM}) generates rules to distinguish the attack behavior from benign applications. The subsequent discussion comprehensively summarizes the details of the proposed algorithm DRsam. Lines 1 to 3 of the algorithm~\ref{DRSAM} correspond to initialization and load phase for each Test Case. Each workload file comprises $p_m$ instances and $q$ microarchitectural features represented by Matrix $\boldsymbol{X}^{(m)}$ and a label vector $\boldsymbol{y}^{(m)}$ containing either `$\boldsymbol{0 \text{ or } 1}$' against each instance of workload $\boldsymbol{W}_m$. Lines 4 and 5 are responsible for independently computing the $\text{Mean}\ \boldsymbol{\mu}^{(m)} \in \mathbb{R}^{1 \times q} $ \ \text{and}\ \text{standard deviation}$\ \boldsymbol{\sigma}^{(m)}\in \mathbb{R}^{1 \times q}$ for each microarchitectural feature in $\boldsymbol{X}^{(m)}$. 
Lines 7 to 9, flag each instance against all microarchitectural features in $\boldsymbol{X}^{(m)}$ using the 3-sigma rule~\cite{3sigma1,3sigma2}. 
\begin{algorithm}[H]
\small
\caption{DRsam (Step 4 in Fig.~\ref{fig:ISCAS})}
\begin{algorithmic}[1]
\label{DRSAM}
\REQUIRE 
Set of workload files against each Test Case 
$\mathbf{W}_1, \mathbf{W}_2, \dots, \mathbf{W}_M \in \mathbb{R}^{p_m \times (q+1)}, \quad m = 1,2,\dots,M $ 
Minimum Support ($\boldsymbol{\theta}_s$) , Minimum Confidence ($\boldsymbol{\theta}_c$)
\EnsureOne   
\ENSURE:  Set of association rules (${\mathcal{R}}$) 
\STATE Initialize ${\mathcal{R}}$ as an empty set 
    \FOR{$m \gets 1$ to $M$}
        \STATE {Load} : \ $ 
\mathbf{X}^{(m)} \in \mathbb{R}^{p_m \times q} \ \ , \ \ \mathbf{y}^{(m)} \in \{0,1\}^{p_m}$
            \FOR{$j \gets 1$ to $q$}
        \STATE $\text{Compute Mean} \  \boldsymbol{\mu}^{(m)} \     \text{and} \ \text{standard deviation} \  \boldsymbol{\sigma}^{(m)} $
         
                \ENDFOR 
        \FOR {each element in $\mathbf{X}^{(m)}$}
        
        \STATE $\mathbf{F}^{(m)} \leftarrow 
        \begin{cases}
        1, & \text{if } {X}^{(m)}_{p,q} > 3.{\sigma}^{(m)} \\
        0, & \text{otherwise}
        \end{cases}$
            \ENDFOR
        
        \FOR{$i \gets 1$ to $p_m$ in $\mathbf{F}^{(m)}$ }
        \STATE Count triggered microarchitectural features$\ t^{(m)}$

         \IF{($y^{(m)}_i = 1 \text{ and } \  \ t_i^{(m)} > q-3)  \ \  \mathbf{OR} \ \ (y^{(m)}_i = 0) $}
            \STATE  keep instances (rows) in $\mathbf{F}^{(m)}$
             \ELSE
           \STATE Discard the instance (row)
             \ENDIF
                \ENDFOR 
    \STATE Concatenate all workloads 
$\boldsymbol{T}_{concat} \ \text{and} \ 
\boldsymbol{Y}_{concat}$

    \ENDFOR
         \FOR{$ \phi = q-3 $ \text{to} $q $ in $\boldsymbol{T}_{concat}$ } 
    \STATE Generate frequent itemsets of size $\phi$ 
    exceeding support threshold $\theta_s$
    \STATE Generate association rules from size $\phi$ itemsets
    satisfying the confidence threshold $\theta_c$
    \STATE Keep rules where support $>$ $\theta_s$ and confidence $>$ $\theta_c$
          \ENDFOR
\STATE Generated rules for \textbf{Test Case 1} are employed to evaluate the performance of this detection method against all other Test Cases
\end{algorithmic}
\end{algorithm}
\vspace{-1em}

\begin{table}[H]
\centering
\caption{Experimental Setup}
\label{tab:Exp-Setup} 
\begin{tabular}{|p{0.35\linewidth}|p{0.55\linewidth}|}
\hline
\textbf{Parameter} & \textbf{Description} \\
\hline
\textbf{Microarchitectural Attacks (RISC-V)} & \textbf{Flush+Fault(Variants)}: Fault and Return \\
\hline
\textbf{gem5 configuration:} \ \ Full System mode &  \textbf{Kernel}: (riscv-bootloader-vmlinux-5.10) , \textbf{Cache hierarchy}:  (Private L1, Public L2), \textbf{ISA}: RISC-V, \textbf{CPU}: O3, 1GHz clock \\
\hline
\textbf{Correalted\ \ \ \ \ \ \ \ \ \ \ \ \ \ \ \ \ \ \ \ \ \  Microarchitectural Features} & L1 I-Cache Misses, Branch Mispredictions, Incorrect conditional branches, Total executed branches, Fetch stalls, TLB accesses, Total Load Instructions, Total Store Instructions \\
\hline
\textbf{rv8 benchmarks} & aes, sha512, norx, dhrystone, miniz, qsort,\ \ \  \ primes \\

\hline
\end{tabular}
\end{table}
\begin{table*} 
\centering
\caption{Comparison of the proposed detection method with the state-of-the-art}

\label{tab:model-comparison} 
\begin{tabularx}{\textwidth}{|c|c|c|c|c|X|X|X|X|c|c|}
\hline
\textbf{Model} & \textbf{Accuracy} & \textbf{Precision} & \textbf{Recall} & \textbf{F1 Score} & \textbf{RG Time/1K } & \textbf{SPP Time/1K } &\textbf{Training Time} & \textbf{Test Time/1K} &\textbf{Benchmarks} & \textbf{New Variants FF} \\
\hline

\textbf{RG} & - & - & - & - & 9.15ms  & 173.3ms  & - & - & - & -\\
\hline

\textbf{Test Case 1}    & 99.89\% & 98.44\% & 100\%  & 99.21\% & -  & 173.3ms  & -  & 155.7ms & NA & NA\\
\hline

Test Case 2 & 99.97\%  & 100\%    & 99.75\%   & 99.87\%    & - & 395ms   & -   & 143.9ms & Yes & Yes \\
\hline
Test Case 3 & 99.96\%    & 100\%    & 99.75\%   & 99.87\%    & - & 542.3ms & - & 155.2ms & Yes & Yes \\
\hline
Test Case 4 & 99.98\%    & 100\%    & 99.75\%   & 99.87\%  & - & 312.7ms  & - & 140.3ms & Yes & Yes \\
\hline
SOTA-a (RF)    & 99\%    & 99\%    & 99\%   & NR & -  & - & NA   & NA  & NA & NA \\
\hline
SOTA-b (SVM)   & 96\%    & 95\%    & 97\%   & NR    & - & -   & NA   & NA & NA & NA\\
\hline
SOTA-c (NB)   & 95\%    & 92\%    & 96\%   & NR  & - & - & NA   & NA & NA & NA \\
\hline
\end{tabularx}
\end{table*}
\vspace{-1em}
According to the 3-sigma rule, if any element in $\boldsymbol{X}^{(m)}$  exceeds more than $3$ standard deviations from the mean value of that particular feature, it is flagged as `1'; otherwise, `0'. This step independently maps each $\mathbf{X}^{(m)} \in \mathbb{R}^{p_m \times q}$ to $\boldsymbol{F}^{(m)}\in \{0,1\}^{p_m\times q}$. Every flagged element of $\boldsymbol{F}^{(m)}$ in any instance 
`$p_m$' is termed as a `triggered feature' if its value is `1'. Lines 10 and 11 count all the triggered microarchitectural features against each instance $p_m$. 
Lines 12 to 16 hold paramount importance by simultaneously capturing the transient behavior during the attack execution and also preventing the `false positives'. From the analysis, the identified set of microarchitectural features are always positively correlated; nevertheless, there are certain transient intervals right before and after the execution of an attack where at least `$q-3$ out of $q$' features get triggered before all `$q$' features are triggered simultaneously. In addition, under the extreme stressed conditions, at most `50 percent' microarchitectural features are triggered. Therefore, for all the workloads labeled as attack `1', this step keeps only those instances against attack workloads where at least `$q-3$ out of $q$' features are triggered. However, for the rv8 benchmark (benign), all the traces are kept as it is. In line 18, all independently processed workload files are concatenated, resulting in $\mathbf{T}_{concat} \in \{0,1\}^{
 K \times q} , \ \mathbf{y}_{concat} \in \{0,1\}^K     \text{where} \ K=\sum_{m=1}^{M} p_m^* \ \text{and} \ p_m^*<p_m $.
Lines 20 to 24 account for rule generation based on predefined support (Definition~\ref{DEF4:Support}) and confidence (Definition~\ref{DEF5:Confidence}) threshold using Apriori. However, apriori is customized to generate the rules only with the antecedents size ($\phi$) between ($q-3 \  to \ q$). The minimum support  and confidence thresholds are set to be $\theta_s = 5 \%  $ and $\theta_c = 90\% $ respectively.
\\ 
\textbf{Step 5: Performance Evaluation}
\\
In this step, the association rules are incorporated to evaluate the performance under different test cases. Accuracy, Precision, Recall, and F1 score are the standard metrics used for the performance evaluation.
\\
\textbf{Step 6: Comparison with State-of-the-Art}
\\
In the final step, we made a detailed comparison of the proposed detection with the state-of-the-art. For the rigorous evaluation of the proposed detection method, the complete rv8 benchmark and all variants of flush+fault attacks are considered at test time.

\section{Results and Discussion} 
This section presents a one-to-one comparison between the proposed detection method and the state-of-the-art. The detailed performance comparison is presented in Table~\ref{tab:model-comparison}. As shown in the \textit{Model} column, the table includes the four test cases described in Table~\ref{tab:TestCases}, along with three state-of-the-art models~\cite{mah3}. The first test case randomly divides 70\% data for association rule mining and 30\% for performance evaluation. In addition, Test Case 1 only considers the fault-based variant of Flush+fault attack without considering any rv8 benchmark workload. Test Case 1 considers the balance class distribution with approximately `5700' attack and benign (operating system) instances, respectively. The rest of the test cases consider all flush+fault variants under different rv8 benchmark workloads. \textit{Accuracy}, \textit{Precision}, \textit{Recall}, and \textit{F1-Score} are standard metrics considered for performance evaluation. The performance comparison reveals a 5.15\% increase in accuracy, 7\% rise in precision, and 3.91\% improvement in recall under the stringent conditions alongside its flexibility to detect new variants of flush+fault attack. The state-of-the-art lacks in both these aspects, i.e., rigorous evaluation under the rv8 benchmarks and its ability to detect new variants of flush+fault attack. Moreover, state-of-the-art don't address the training and test times for attack detection. To address this gap, we also reported execution time for association rules generation (RG Time), statistical preprocessing (SPP Time), and testing to demonstrate the efficiency of the proposed detection method. To standardize evaluation across all test cases, execution time is reported with respect to a reference of 1K samples. The future research direction aims to evaluate the effectiveness of the proposed detection method against a wide range of microarchitectural attacks and introduce efficient mitigation solutions.
\vspace{-0.5em}
\vspace{-0.5em}
\begin{table} [H]
\caption{Test Cases for Performance Evaluation}
\label{tab:TestCases}
\centering
\begin{tabularx}{\linewidth}{|c|X|X|}
\hline
\textbf{Test Cases} & \textbf{Benchmarks} & \textbf{Flush+Fault Variants} \\
\hline
Test Case 1 & Only Operating System & Fault-based variant \\
\hline
Test Case 2 & aes, sha512, dhrystone, norx & Fault and Return variants \\
\hline
Test Case 3 & qsort, prime, miniz & Fault and Return variants \\
\hline
Test Case 4 & aes, sha512, dhrystone, norx, qsort, prime, miniz & Fault and Return variants \\
\hline
\end{tabularx}
\end{table}

\section{Conclusion}
In this paper, we presented a robust method for detecting flush+fault microarchitectural attacks in RISC-V. The proposed detection method (DRsam) leverages statistical preprocessing and association rule mining. The effectiveness of the proposed detection method is evaluated under the rigorous and stringent conditions by incorporating the complete rv8 benchmark and all variants of flush+fault attacks. A comprehensive comparison with the state-of-the-art demonstrates enhanced interpretability,  5.15\% increase in accuracy, 7\% rise in precision, and 3.91\% improvement in recall under the stringent conditions alongside its flexibility to detect new variants of flush+fault attack.
\clearpage
\section*{Acknowledgment}
This work was supported by the Estonian Research Council grants PSG837.

\bibliography{ISCAS.bib}         




\end{document}